\documentstyle[prl,aps,multicol,epsf]{revtex}

\title{Force--Velocity Relations of a Two--State Crossbridge\\ Model
for Molecular Motors}

\author{Andrej Vilfan, Erwin Frey, and Franz Schwabl}

\address{Institut f\"ur Theoretische Physik, Technische Universit\"at
  M\"unchen, D--85747 Garching, Germany}

\date{April 21, 1998}

\begin{document}
\maketitle
\pacs{PACS numbers: 05.40.+j, 87.10+e, 87.22Jb}

\begin{abstract}
We discuss the force-velocity relations obtained in a two-state crossbridge
model for molecular motors. They can be calculated analytically in two limiting
cases: for a large number and for one pair of motors. The effect of the
strain-dependent detachment rate on the motor characteristics is studied. It
can lead to linear, myosin-like, kinesin-like and anomalous curves. In
particular, we specify the conditions under which oscillatory behavior may be
found.
\end{abstract}

\begin{multicols}{2}
\narrowtext Understanding the molecular mechanism underlying biological motors
has recently attracted increasing interest in biology as well as physics
\cite{block96}.  Motor proteins such as myosin, kinesin and dynein moving along
molecular tracks are involved in a wide range of processes essential for life,
{\em e.g.}\ cell division, muscle contraction, and intracellular transport of
organelles.  For many decades exclusively data from physiological measurements
on muscles \cite{hill38} provided experimental information for modeling
molecular motors \cite{huxley57,huxley71}.  In recent years, a variety of {\it
in vitro\/} techniques allowed the observation of single motor proteins
\cite{simmons96} and gave new insights into the basic principles underlying
their operation.  Not only new theoretical models for single-molecule motors
\cite{ajdari92,peskin95,duke96,derenyi96} were inspired by these experiments,
but also new models for cooperative motors
\cite{leibler93,prost95,juelicher97b}.

The theoretical models can follow two different goals. Either they are designed
to fit as many physiological experiments as possible by including many (up to
six) different states, or one uses simplified models (mostly with two states)
in order to extract the generic features of motion generation and classify the
motors according to their properties \cite{leibler93,juelicher97b}.  Latter
models fall into two classes, one using a specific conformational change (power
stroke) in the motor molecule \cite{leibler93}, the other a ratchet mechanism
\cite{juelicher97b}.  A striking result of the ratchet models was the
prediction of spontaneous oscillations of cooperative motors
\cite{juelicher97}, which might explain the oscillatory behavior of muscles
\cite{yasuda96}.

Here we discuss the force-velocity relations of a two-state model with
strain-dependent detachment rates. Depending on the functional form of these
rates, the model can show a much greater variety of phenomena than previously
discussed \cite{huxley57,leibler93}.  These include linear, hyperbolic,
anomalous or kinesin-like force-velocity relations.  In the two-state model
each motor molecule has two long living states: attached and detached. This
corresponds to the model described by Leibler and Huse \cite{leibler93} when
only the time limiting steps important for mechanical properties are taken into
account.  Two-state models have also been used previously for myosin
\cite{huxley57} and kinesin \cite{peskin95,duke96} as well as in ratchet models
\cite{prost95,juelicher97b,astumian96}.  We generalize the two-state model by
introducing arbitrarily strain-dependent transition rates and discrete binding
sites.  Both extensions are crucial for a qualitative and quantitative
explanation of experiments.

\begin{figure}
\centerline{\epsfxsize=0.5\columnwidth \epsffile{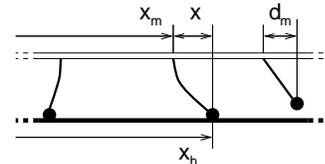}}
\caption{\label{fig1} 
Schematic model for the motor heads running along a molecular track.}
\end{figure}

The {\em model} is defined as follows. Let $x_h$ denote the position of a (free
or bound) head and $x_m$ the position of its root at the backbone, as sketched
in Fig.\ \ref{fig1}.  Deformations of one head can then be described in terms
of a harmonic potential ${\cal H}=U(x_h-x_m-x_d)$ with $U(x)\equiv \frac12 k_m
x^2$.  After attaching to or detaching from the fiber, a conformational change
in the head, described by shifting the potential by the distance $d_m$, takes
place, {\em i.e.}\ $x_d=0$ in the attached (A) state and $x_d=d_m$ in the
detached (D) state. This is the {\em first spatial asymmetry} in our model and
constitutes the basic mechanism for the generation of directed motion. We
assume that the transitions between the two states occur stochastically with
characteristic times $t_a$ and $t_d$.  While the attachment rate $t_d^{-1}$ can
be assumed to be constant, there is strong experimental evidence
\cite{finer94,fenn24} for a strain-dependent detachment rate $t_a^{-1}$ of
myosin.

We will show that different functions $t_a(x)$ describing strain-dependence of
the detachment rate leads to various interesting phenomena, which are the main
topic of our discussion. In generally $t_a(x)$ will be an asymmetric function,
thus bringing a {\em second asymmetry} into the model.  The binding sites are
discrete with a separation of $a=5.5\,\mbox{nm}$ on actin ($8\,\mbox{nm}$ on
tubulin \cite{svoboda93}). Before binding to a site, the head position
fluctuates due to thermal motion. We assume that even in the affine state there
is a time scale separation between the diffusion and the attachment time of the
free head. This leads to the probability that a head positioned at $x_0$ binds
to site $i$ given by the Boltzmann weight $W_i\propto{\exp[-\beta U(-x_0+i
a)]}$ \cite{note_weight}. An experimental estimate for the amplitude of thermal
fluctuations of a free myosin head with data from Ref.\ \cite{finer95}
($k_m=0.4\,{\rm pN/nm}$, $d_m=10\,{\rm nm}$) gives $\sigma=\sqrt{k_{\rm
B}T/k_m}\approx 0.3\,d_m$.

We start our analysis by considering a large group of $N$ rigidly coupled
independent motors, a situation typical for the actin-myosin motor in muscles.
Then the fluctuations resulting from the stochastic operation of single motors
can be neglected. We set up a Master equation for the probability densities
$\Phi_a(x,t)$ for a motor being attached at $x$ at time $t$ and $\Phi_d(t)$ for
a motor being detached. We also need the probability density $P(x,x_m)$ that a
motor at $x_m$ attaches at a distance $x$ from its root. This will depend on
the actual position of the motor head $x_m$ with respect to the binding sites
($x_i = i a$). But, since generically myosin remains in the detached state for
a relatively long time we may assume the root position before attachment to be
completely random with respect to the discrete binding sites and replace
$P(x,x_m)$ by its average $P(x) = \int_0^a dx_m P(x,x_m) /a$.  Then, $P(x)$
becomes
\begin{equation}
\label{eq1}
P(x)=\frac 1a \frac{e^{-\beta
U(x-d_m)}}{\sum_j e^{-\beta U(x-d_m+ja)}} \, ,
\end{equation}
and the probability densities $\Phi_a$ and $\Phi_d$ obey the following Master
equations
\begin{eqnarray}
\label{eq2}
\left(\partial_t- v \partial_x\right)\Phi_a(x,t)&=&
\frac{\Phi_d(t)}{t_d} P(x) 
-\frac{\Phi_a(x,t)}{t_a(x)} \nonumber\\
\partial_t\Phi_d(t)&=& -\frac {\Phi_d(t)}{t_d} +\int \! dx \,
 \frac{\Phi_a(x,t)}{t_a(x)} \, ,
\end{eqnarray}
with normalization $\Phi_d(t)+\int \Phi_a(x,t)dx=N$. The force produced by the
group of motors is given by
$F(t)=\int dx\,\Phi_a(x,t)\partial_x U(x)$.

\begin{figure}
\begin{center}
\begin{tabular}{ll}
{\epsfysize=0.3\columnwidth\epsffile{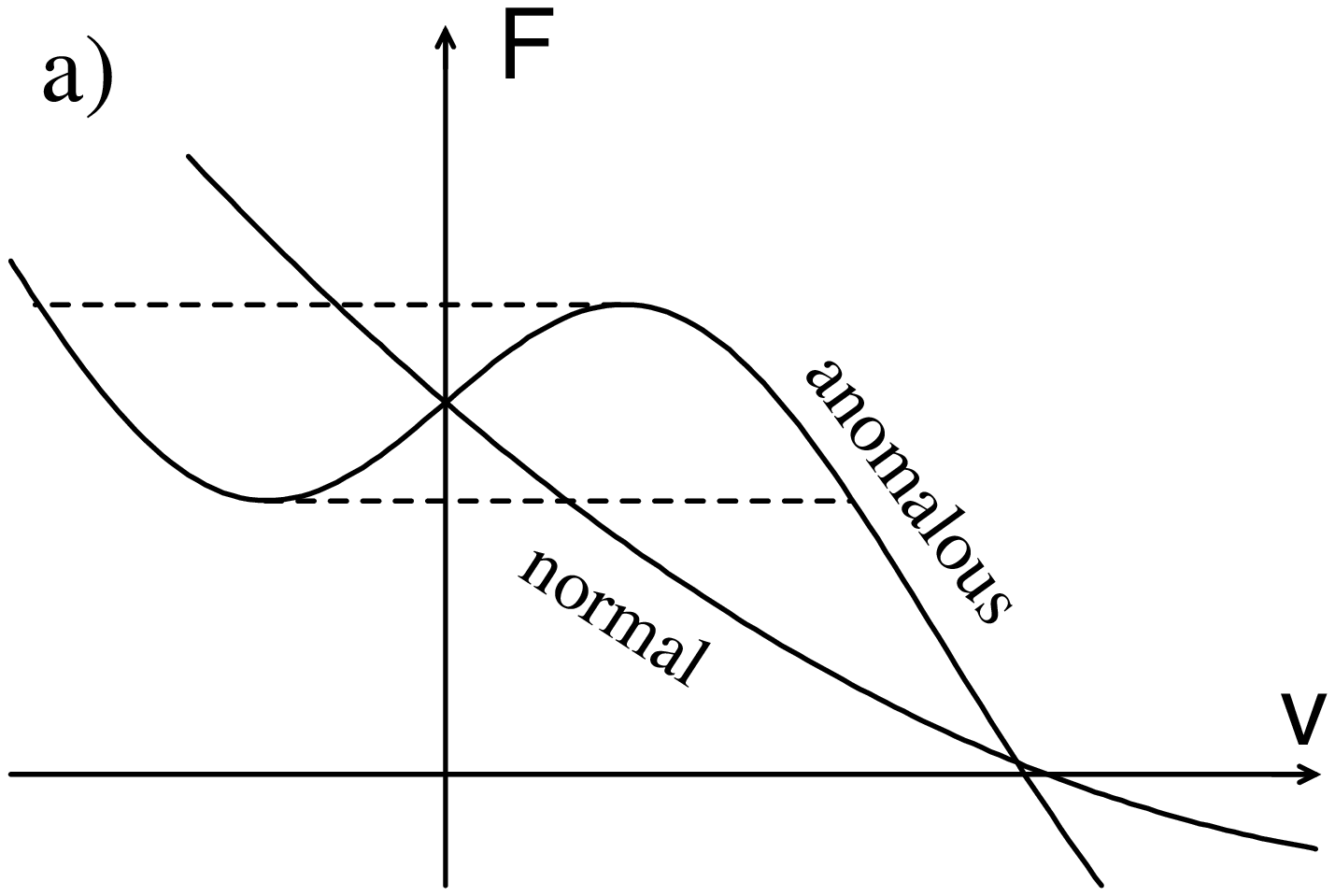}}
&
{\epsfysize=0.3\columnwidth\epsffile{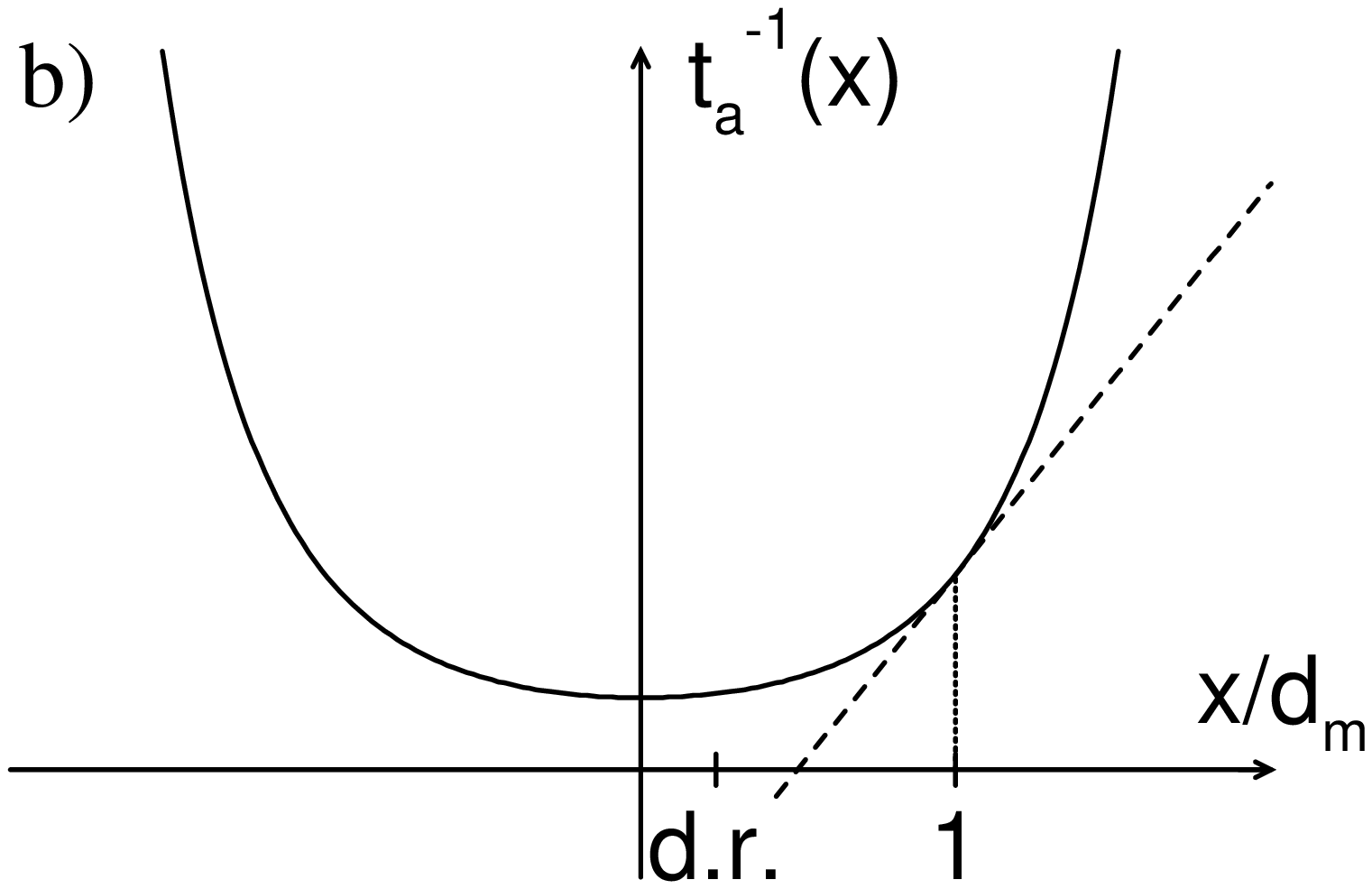}}
\end{tabular}
\end{center}
\caption{\label{fig2}
{\bf a)} Examples of a normal an an anomalous $F-v$ relation, leading to
oscillating behavior. {\bf b)} A graphical representation of the criterion for
the occurance of oscillations. If the tangent to the detachment rate as a
function of $x/d_m$ in the point 1 crosses the $x$-axis right of the point
given by the duty ratio (d.r.) at zero velocity (as shown above), the
force-velocity relation is anomalous with a hysteresis around $v=0$.}
\end{figure}

For a constant positive velocity we have to find stationary solutions of Eq.\
\ref{eq2}:
\begin{eqnarray}
\label{eq3}
\Phi_a(x)&=&\frac{N \int dy\,G(x,y)\,P(y)}{v t_d+\int
dx' \int dy\,G(x',y)\,P(y)} \nonumber\\
G(x,y)&=&\exp\left(-\int_x^y\frac{dx'}{v t_a(x')}\right) \theta(y-x)\,.
\end{eqnarray}
$G(x,y)$ is a Green's function which can be interpreted as the probability that
a motor which got bound to the fiber at position $y$ still remains bound when
its position reaches $x$.

For a harmonic potential $U(x)$ and a strain-independent detachment rate one
gets a linear force-velocity relation $F/N=t_a k_m (d_m-vt_a)/(t_a+t_d)$.  It
neither depends on temperature nor on the distance between the binding sites.
More complex functions $t_a(x)$ of course lead to other forms of the
force-velocity relations.  They can be classified into two groups: the normal
ones with a monotonously decreasing force for an increasing velocity and the
anomalous ones, showing hysteretic behavior (Fig.\ \ref{fig2}a).  The reason
why anomalous relations are interesting is that they allow two different
velocities for the same external force. For instance, in a harmonic external
potential this leads to spontaneous oscillations if the hysteresis spreads over
$v=0$. Such oscillations were first proposed in a two state ratchet model by
J\"ulicher and Prost \cite{prost95,juelicher97,juelicheres
97b}. Here we show how
such a mechanism can be implemented in crossbridge model.

Upon neglecting the discreteness of the binding sites and thermal fluctuations,
a simple {\it sufficient} algebraic criterion for the occurance of these
oscillations can be derived. The zero velocity point certainly lies in a
hysteretic range if the slope of the force-velocity relation is positive there.
Due to the simplification mentioned above we set $P(y)=\delta(y-d_m)$ in Eq.\ 
\ref{eq3} and calculate the derivative
\begin{eqnarray}
\left.\frac{dF}{dv}\right|_{v=0}&=&N\frac{k_m t_a^2(d_m)}{t_d+t_a(d_m)}\Bigl(
-1\nonumber\\
&&+\frac{t_d}{t_d+t_a(d_m)}d_m t_a(d_m) \frac d
{dx}\left. t_a^{-1}(x)\right|_{x=d_m}\Bigr)\;.
\end{eqnarray}
If its value is positive, the force-velocity relation certainly shows anomalous
behavior. A graphical representation of this criterion is shown in Fig.\
\ref{fig2}. 

\begin{figure}
\centerline{\epsfxsize=0.7\columnwidth \epsffile{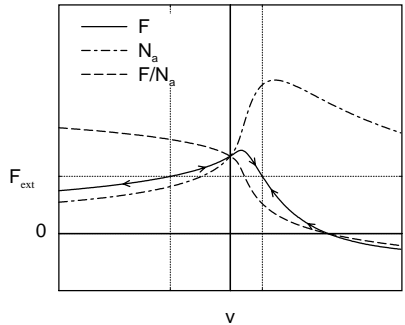}}
\caption{\label{fig3}
An example of the $F-v$ curve for\protect\linebreak $t_a(x)\propto \exp(-2 \left| x
\right| / d_m)$ and a low duty ratio. The solid line shows the mean force
motors would produce at a given velocity. The dashed line shows the mean number
of attached motors and the dash-dotted one the mean force per attached motor.
Seeking for the velocity at a given force one obtains a stable (higher $v$) and
an unstable (lower $v$). A third solution is always $v\to -\infty$, but to
obtain it one has to include an infinitesimally small friction term.}
\end{figure}

The force-velocity relation as calculated by now describes the mean force a
group of motors produces when moving with a given constant velocity. However,
the situation is usually reversed and one is interested in the mean velocity at
a constant force. Of course, both situations are equivalent in the limit of
large $N$. But for a finite $N$ the motion actually occurs stepwise. This
raises the question, how the motors remember on which limb of the hysteresis
they currently move.  The quantity that actually distinguishes between both
limbs is the number of currently attached motors. Together with the external
force it uniquely defines the velocity. This follows from the fact that the
force per attached motor decreases monotonously with increasing velocity (Fig.\
\ref{fig3}), as can be seen from Eq.\ \ref{eq3}.  In a finite system the
number of motors fluctuates and if it passes a threshold value the velocity
jumps from one stable state into the other.  The probability for such jumps is
highest if the original state is close to the edge of the hysteresis and the
number of motors low.

An example of a function always leading to a {\em normal} $F-v$ relation is
$t_a(x)=\exp(\alpha x)$. It states that the lifetime of the attached state is
larger for those heads that have just gone through the power stroke and produce
maximum force than for those which have already done their work and now pull
backwards.  As a consequence the duty ratio becomes lower at higher velocities.
This idea has already been used by A.F. Huxley \cite{huxley57}.  Such a
dependence is needed for an explanation of the approximately hyperbolic
force-velocity dependence in muscle systems.  Physiological data by Hill
\cite{hill38} are perfectly fitted within the above analytic results by
choosing $\alpha\,d_m=0.55$ and $t_d\gg t_a^0$.  Quantized binding sites and
thermal fluctuations are found to play only a minor role (their neglect leads
to almost the same curve with $\alpha\,d_m=0.58$).

A function that can lead to an {\em anomalous} $F-v$ relation is $t_a(x)\propto
\exp(-2 \left| x \right| / d_m)$, as shown in Fig.\ \ref{fig3}.  For a
sufficiently low duty ratio the point $v=0$ lies within a hysteresis and is
unstable if the force is held constant. Instead, a positive finite solution is
possible or a negative with $v\to -\infty$. To obtain the latter from our
equations, an infinitesimally small friction term has to be added.  Now if one
lets such a group pull against a harmonic spring with the other end fixed, the
extension of the spring oscillates in a sawtooth-like manner with flat
ascending slopes (motors working against the spring force) and very steep
descending slopes (the spring force pulling motors backwards). Very similar
behavior has been observed in muscles under some conditions (including low
${\rm Ca}^{2+}$ concentration, which indeed means a low duty-ratio)
\cite{yasuda96}.  However, other explanations which suppose that the
oscillations are induced by the regulatory system are possible as well
\cite{smith94}.

So far, our discussion has focused on situations where motors are operating in
large groups. There is, however, a second scenario, where only a few molecular
motors cooperate at a time, {\em e.g.}\ when kinesin transports vesicles along
microtubules. Modeling them is guided by the following experimental
observations: A kinesin molecule with two heads can move over long distances
without detaching from the microtubule \cite{svoboda94a}. Although it is not
yet completely clear how the two kinesin heads ``walk'' along the
protofilaments \cite{block95}, there are good arguments to use a model with
$8\,{\rm nm}$ periodicity \cite{howard96} where each binding site can be
occupied only with one head at a time.  Single headed kinesin can move
microtubules if cooperating in larger groups, but not as fast as double-headed
\cite{vale96}. The velocity decreases with increasing load almost linearly
\cite{svoboda94a,meyhoefer95}. Above the stall force kinesin shows
back-and-forth movement, but does not walk backwards \cite{coppin97}.  Forward
loads can increase the velocity many times \cite{coppin97}.

From the fluctuation analysis \cite{svoboda94b,schnitzer97} it is quite certain
that both the process of attachment and of detachment include an asymmetry
between the forward and the backward direction (a new head attaches in front of
the other one and the rear head detaches more probably than the front one).
The low variance \cite{note_variance} ($r\approx 0.5$) is not {\it a priori}
inconsistent with a model where only one of both symmetries is present ({\em
e.g.}\ one where steps with $8\,{\rm nm}$ and $0\,{\rm nm}$ occur with equal
probabilities), but then everything else in the duty cycle including the dwell
times would have to be completely deterministic, which does not seem realistic.
The behavior at superstall forces \cite{coppin97} additionally implies that one
of these both asymmetries remains over the whole force range, while the other
one reverses at higher forces.  In our discussion we restrict ourselves to a
model in which the attachment asymmetry gets reversed with increasing load
while the detachment asymmetry remains. This approach has already been used by
Peskin and Oster \cite{peskin95} and in a similar way by Duke and Leibler
\cite{duke96}. This, however, does not mean that we consider the other case
less realistic.

\begin{figure}
\centerline{\epsfxsize=0.7\columnwidth \epsffile{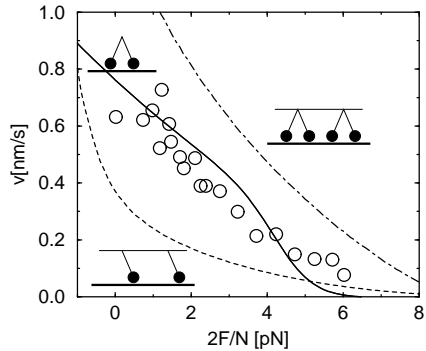}}
\caption{\label{fig4}
Force-velocity curves for kinesin. Experimental data from Ref.\
\protect\cite{svoboda94a} at saturating ATP concentration and the theoretical
curve for $\alpha=0.5\,{\rm nm}^{-1}$, $t_d=3\,{\rm ms}$, $t_a^0=700\,{\rm
ms}$, $d_m=4\,{\rm nm}$ and $k_m=1.1\,{\rm pN}/{\rm nm}$. The middle curve
shows the result for one double-headed molecule, the upper one for many coupled
double-headed molecules and the lower one for many single-headed molecules.}
\end{figure}

The central result is again the force-velocity relation. Because the velocity
is not temporally constant as for $N\to \infty$, it has to be calculated
directly from transition rates for a constant force.  Since an attachment two
sites away from the other head is very unprobable, we take only the attachment
rate at the front (f) or rear (r) side of the other head into account:
$R_a^{f/r} = {\cal N} \exp[ \pm \beta k_m a (d_m-F/k_m)/2]$; ${\cal N}$ is
chosen such that both rates add up to $t_d^{-1}$.  The respective detachment
rates are $R_d^{f/r} = t_a^{-1}((\pm a +F/k_m)/2)$.  This gives
\begin{equation}
v(F)=\frac{a/2}{t_d+1/(R_d^f+R_d^r)}\left(\frac{R_d^r-R_d^f}{R_d^r+R_d^f}+
\frac{R_a^f-R_a^r}{R_a^f+R_a^r}\right)\;.
\end{equation}
The linear force-velocity curve has led some authors \cite{block95,svoboda94a}
to the conclusion that there is no strain-dependence of the detachment rates.
This conclusion, however, is only valid in a model with continuous binding
sites \cite{leibler93}. Taking into account discrete binding sites actually
leads to a nonlinear, S-shaped curve if the detachment rates are
strain-independent. Again the simplest choice is $t_a(x)=\exp(\alpha x)$. Using
a reasonable set of parameters the model is able to reproduce the nearly linear
dependence reported in Refs.\ \cite{svoboda94a,meyhoefer95} (Fig.\ \ref{fig4})
with extensions similar to those in Ref.\ \cite{coppin97}. Fig.\ \ref{fig4}
further shows the velocity for a large number of coupled double-headed and
single-headed kinesin molecules. When comparing them to experiments care has to
be taken since the pairs are in reality coupled elastically to the backbone,
which leads to lower velocities. Beside that both curves depend very
sensitively on the choice of $t_a(x)$. Nevertheless, they show clearly that the
``repulsion'' between heads already causes a significant velocity difference
between single- and double-headed kinesin.

Another quantity of interest is the probability for the whole molecule
detaching from the microtubule during one step is $P_L=t_d/t_a(F/k_m)$,
yielding $5\%$ at zero load and saturating ATP concentration, somewhat higher
than comparable observations ($1.3\%$) \cite{vale96}.

In summary, we have shown that a generalized two-state crossbridge model for
molecular motors can lead to a much larger variety of phenomena than previously
discussed. We have found analytical results in two limiting cases: for a large
number of rigidly coupled motors and for one pair. In the first case we show
how different functions describing the strain-dependence of the detachment rate
result in linear, hyperbolic or even anomalous force-velocity relations and
give a simple algebraic criterion for the latter.  Discrete binding sites play
only a minor role. For one pair of motors force-velocity-relations as measured
on kinesin can be reproduced. They depend crucially on the displacement between
the binding sites. The model also shows a significant difference between
single- and double-headed kinesin when operating in large grups.

\section*{Acknowledgments}

We are grateful to Klaus Kroy, Rudolf Merkel and Erich Sackmann for helpful
discussions. This work has been supported by the Deutsche
Forschungsgemeinschaft under contract no.\ SFB 413.  A.V. and E.F. would like
to acknowledge support from the Cusanuswerk and by a Heisenberg fellowship from
the Deutsche Forschungsgemeinschaft, respectively.

\end{multicols}


\begin{thebibliography}{10}
\bibitem{note_variance}
An important feature of the stochastic trajectory is the variance of
  displacement \cite{svoboda94b,schnitzer97}. The quantity $r={\frac{d}{dt}
  \langle (x(t)\!-\!\langle x(t)\rangle)^2 \rangle / \langle v \rangle a}$ is a
  measure for the randomness of the movement. Its value is $1$ for a Poisson
  stepper and $0$ for a deterministic process. A two-state model can reach a
  minimum of $r=0.5$. The parameters used above lead to $r=0.62$ at zero load,
  which is not far from the experimental value $r=0.52\pm0.2$
  \cite{svoboda94b,schnitzer97}.
\end{thebibliography}
\end{document}